\documentstyle[11pt,aaspp]{article}
\newcommand{\postscript}[2]
 {\setlength{\epsfxsize}{#2\hsize}
  \centerline{\epsfbox{#1}}}
\def\ref#1{\par\noindent \hangindent=0.4in \hangafter=1 #1 \par}
\def\eqalign#1{\null\,\vcenter{\openup\jot \m@th
  \ialign{\strut\hfill$\displaystyle{##}$&$
     \displaystyle{{}##}$\hfill \crcr#1\crcr}}\,}
\def\tempest%
{\begin{array}{ccc}
1 & 1 & 1 \\
1 & 1 & 1 \\
4 & 3 & 8
\end{array}}

\begin{document}

\title{Einstein Radii from Binary Lensing Events${}^{1}$}
 
\author{Cheongho Han}
\author{Andrew Gould$^2$}
\affil{Ohio State University, Department of Astronomy, Columbus, OH 43210}
\affil{e-mail: cheongho@payne.mps.ohio-state.edu}
\affil{e-mail: gould@payne.mps.ohio-state.edu}
\footnotetext[1]{submitted to {\it Astrophysical Journal}, Preprint:
OSU-TA-11/96}
\footnotetext[2]{Alfred P.\ Sloan Foundation Fellow}
 
\begin{abstract}

We show that
the Einstein ring radius and transverse speed of a lens 
projected on the source plane, $\hat{r}_{\rm e}$ and  $\hat{v}$,
can be determined from the light curve of a binary-source event,
followed by the spectroscopic determination of the orbital elements 
of the source stars.
The determination makes use of the same principle that allows one 
to measure the Einstein ring radii from finite-source effects.
For the case when the orbital period of the source stars is much longer 
than the Einstein time scale, $P\gg t_{\rm e}$,
there exists a single two-fold degeneracy in determining 
$\hat{r}_{\rm e}$.
However, when $P \lesssim t_{\rm e}$ the degeneracy can often be broken
by making use of the binary-source system's orbital motion.
%Once $\hat{r}_{\rm e}$, and thus $\hat{v}$ are determined, one can 
%distinguish self-lensing events in the Large Magellanic Cloud 
%from Galactic halo events.
For an identifiable 8\% of all lensing events seen toward the Large 
Magellanic Cloud (LMC), one can unambiguously determine whether the 
lenses are Galactic, or whether they lie in the LMC itself.
The required observations can be made after the event is over and 
could be carried out for the $\sim 8$ events seen by Alcock et al.\ and 
Aubourg et al..
In addition, we propose to include eclipsing binaries as sources for
gravitational lensing experiments.

\end{abstract}


\newpage
\section{Introduction}

There are many different effects that make a light curve of 
a microlensing event 
deviate from its characteristic achromatic and symmetric form:
luminous lenses (Kamionkowski 1995; Buchalter, Kamionkowski, \& Rich 1995),
differential magnification during close encounters (Gould 1994;
Nemiroff \& Wickramasinghe 1994; Witt \& Mao 1994; Witt 1995;
Loeb \& Sasselov 1995; Gould \& Welch 1996),
parallax effects caused by the Earth's orbital motion
(Gould 1992; Alcock et al.\ 1995), and finally binary-lens events 
(Mao et al.\ 1994; Axerlod et al.\ 1994; Udalski et al.\ 1994;
Mao \& Di Stefano 1995; Alard, Mao, \& Guibert 1995; Alcock et al.\ 1996c).
Whenever any of these distortions is detected,
it provides information about the physical parameters
of individual lenses: distance to the lens for luminous lens,
lens proper motion, $\mu=v/D_{\rm ol}$, for differential magnification,
observer-plane projected Einstein ring radius,
$\tilde{r}_{\rm e} = (D_{\rm os}/D_{\rm ls})r_{\rm e}$, for parallax,
and the geometry of a lens binary system and sometimes the proper motion 
for binary lens events.
Here, $v$ is the speed of the lens relative to the Earth-source line 
of sight, and the physical and angular Einstein ring radius are related 
to the physical parameters of the lens by
$$
r_{\rm e} = \left( {4GM_{L} \over c^2}{D_{\rm ol}D_{\rm ls} 
\over D_{\rm os}} \right)^{1/2},\qquad
\theta_{\rm e} = {r_{\rm e} \over D_{\rm ol}},
\eqno(1.1)
$$
where  $D_{\rm ol}$, $D_{\rm ls}$, $D_{\rm os}$ are the distances
between the observer, source, and lens, $r_{\rm e}$ is the
physical size of the Einstein ring, and $M_{L}$ is the
mass of the lens.

The light curve can be also distorted when the source is composed 
of a binary system: binary-source event (Griest \& Hu 1992).
The binary-source event light curve distortions take various forms
depending on many factors, e.g., the angular size of the projected
separation between the source stars, source trajectories within the
Einstein ring, the angular size of the Einstein ring projected
onto the source plane, and the orbital motion.
Griest \& Hu (1992) concentrated on binary-source events for 
which the binary period is long compared to the event time scale.
They briefly discussed the possibility of events with short-period
binary sources and illustrated the dramatic oscillation which
these could in principle generate.
They noted, however, that for the expected lens parameters the amplitude
of these oscillations would be extremely small.

In the present paper, by contrast, we concentrate on binary-source
events where the binary stars move substantially, i.e., short-period 
binaries.
We show that the Einstein ring radius
and the transverse speed projected on the source plane,
$\hat{r}_{\rm e}$ (`source-plane Einstein ring radius')
and $\hat{v}$ (`source-plane transverse speed'),
can be determined from the light curve of a binary-source event,
provided that the observations are followed by spectroscopic 
determination of the binary-source orbital elements.
Throughout this paper, we use  a ``hat'' ($\hat{\ }$) to represent
a quantity projected onto the source plane.
The source-plane Einstein radius and transverse speed are defined by
$$
\hat{r}_{\rm e} = D_{\rm os}\theta_{\rm e} =
r_{\rm e}{D_{\rm os}\over D_{\rm ol}}, \ \
\hat{v} = {\hat{r}_{\rm e} \over t_{\rm e}},
\eqno(1.2)
$$
where $t_{\rm e}=r_{\rm e}/v$ is the Einstein ring crossing time.
Note that when the source distance is known [e.g., for observations 
toward the Large Magellanic Cloud (LMC)],
measuring $\hat{r}_{\rm e}$ and $\hat{v}$ is equivalent to measuring
$\theta_{\rm e}$ and the proper motion $\mu$ since
$\theta_{\rm e} = \hat{r}_{\rm e}/D_{\rm os}$ and
$\mu = \hat{v}/D_{\rm os}$.
The basic principle that makes this measurement possible is that a
binary acts like an enormous finite source and therefore
is much more susceptible to finite-source effects than are single stars.
Once the proper motion is measured, one can uniquely separate
Galactic versus LMC self-lensing events because of the large
difference in the proper motions between the two populations of 
events (see \S\ 6).

Short-period binary sources are important for several reasons.
First, it is these events that allow one to unambiguously measure
the proper motion.
For long period binaries, one can determine $\hat{r}_{\rm e}$, but
with a two-fold degeneracy (see \S\ 3).
Second, for lenses in the LMC the amplitude of the oscillations
in the flux due to the binary-source effect
is expected to be of order 10\% and would be easily observable 
(see \S\ 5).
Within the framework of the standard model, LMC events are expected to
be relatively rare.
However, Sahu (1994) has argued that essentially all the events
currently detected toward the LMC are due to LMC lenses.
This may seem unlikely in view of the large optical depth
(Alcock et al.\ 1996a), but it is nevertheless
important to test this hypothesis.
Since the alternative is that the halo is composed in large part
of Massive Compact Halo Objects (MACHOs).
As we show in \S\ 5, about 10\% of LMC sources are short-period binaries.
These binaries allow a direct test of the Sahu (1994) hypothesis.
Third, while the effects are smaller for  Galactic lenses,
they are not negligible.
The significant advances now being made in rapid detection and 
follow-up observations (Pratt et al.\ 1995; Albrow et al.\ 1996;
Alcock et al.\ 1996b) open the
possibility that even 1\% or 2\% oscillations caused by 
binary sources may soon be measurable.
Finally, binary sources are competitive with other methods of measuring
proper motion toward the Galactic bulge.

\section{Proper Motion from a Binary-Source Event}

The light curve of a binary-source event is the sum of  
light curves of individual sources
and is represented by
$$
F = \sum_{j=1}^{2} A_{j}F_{0,j};\ \ \
A_j = {u_j^2 + 2 \over u_j(u_j^2+4)^{1/2} },
\eqno(2.1)
$$
where $j=1,2$ denote the primary and secondary
source stars, $F_{0,j}$ are the unmagnified fluxes,
and $u_j$ are the projected locations of the source stars
with respect to the lens in units of $r_{\rm e}$.
%For a slowly orbiting binary system, the source stars move along
%a straight line due to the source-lens transverse motion.
%On the other hand, for a fast-orbiting binary, the positions of source
%stars are determined from the combination
%of the linear transverse and orbital motions (see \S\ 2 for details).

When stars in a binary are have negligible orbital motion, 
the source stars move along a straight line due to the 
source-lens-observer transverse motion (case I).
Then the source positions are well approximated by
$$
u_j^2 = \omega^2 (t-t_{0,j})^2 + \beta_j^2,
\eqno(2.2)
$$
where $\omega=t_{\rm e}^{-1}$, $\beta_j$ are the impact parameters,
and $t_{0,j}$ are the times of maximum magnification.
Note that the Einstein time scale is same for both sources.
Then the projected (two-dimensional) separation between two stars 
is related to
$\hat{r}_{\rm e}$ by
$$
\ell = \ell_1 + \ell_2 =
\hat{r}_{{\rm e}\pm} [(\omega \Delta t_0)^2 +
\Delta\beta_{\pm}^2]^{1/2},
\eqno(2.3)
$$
where 
the coordinates $(x,y)$ are
respectively parallel and perpendicular to the
direction of the lens motion and 
$\ell_1=\ell/({\cal Q}_{\rm M}+1)$ and 
$\ell_2={\cal Q}_{\rm M}\ell_1$ are the 
separations between the center of mass (CM) and individual sources.
Here ${\cal Q}_{\rm M}=M_1/M_2$ is the mass ratio between the 
source stars.
If $\ell$ can be determined (see below), is also possible to determine
$\hat{r}_{\rm e}$, provided one can decompose the light curve into its
individual components and so measure $\Delta t = |t_{0,1}-t_{0,2}|$
and $\Delta \beta _{\pm} = |\beta_1 \pm \beta_2 |$.
In \S\ 3, we discuss the twofold degeneracy in $\hat{r}_{\rm e}$ 
induced by the ambiguity in the impact parameter difference, 
$\Delta\beta_{\pm}$.

On the other hand, when the orbital motion of a binary is important
(case II), 
the positions of the source stars are the combination of the 
linear transverse motion of the CM and the 
orbital motion of component stars around the CM;
$$
%{\bf u}_{j}(t) = {\bf u}_{\rm CM} + \delta {\bf u}_{j} (t),
u_{j}^{2} = 
u_{\rm CM}^2
-2u_{\rm CM}\left( {\ell_j\over \hat{r}_{\rm e}} \right) \cos \theta_j
+\left( {\ell_j\over \hat{r}_{\rm e}} \right)^2,
\eqno(2.4)
$$
where $\theta_{j}$ is the lens-CM-source angle.
The values of $u_j$ and $u_{\rm CM}$ are determined from the light curve.
Then, once the orbital motion, i.e., $\ell_j [\theta_{j}(t)]$, 
is known, one can determine $\hat{r}_{\rm e}$.

The projected separation between source stars is related to 
the intrinsic (3-dimensional) separation $\ell_0$ by
$$
\ell =
%\cases{
%\ell_0,
%&  $\sin \Omega = 0$, \cr
\ell_0 \left\vert {\sin (\Theta +\psi_{\rm P}) \over
\sin \Omega } \right\vert \cos i,
%& otherwise, \cr
%}
\eqno(2.5)
$$
where $\Theta$ is the true anomaly, $\Omega$ is the position angle 
to a star measured from the ascending node, 
$i$ is the inclination angle, and $\psi_{\rm P}$ is the longitude 
of the periastron measured from the ascending node. 
Here the position angle is measured along the projected orbital plane, 
while the longitude is measured along the true orbital plane.
The intrinsic separation between the stars is given by
$$
\ell_0(\Theta ) =
{a (1-\epsilon^2 ) \over 1 + \epsilon \cos \Theta }.
\eqno(2.6)
$$
Here $\epsilon = (1-b^2/a^2)^{1/2}$ is the eccentricity, 
where $a$ and $b$ are the
semimajor and semiminor axes of the orbit.
When the orbit is seen face-on ($i=0^{\circ}$), 
$\Omega = \Theta + \psi_{\rm p}$, and thus 
$\ell = \ell_0$.

For a gravitationally lensed binary-source system, the 
individual masses can be estimated to a first approximation 
from the luminosities and colors.
One can then further constrain the stellar masses
by determining the stellar types from follow-up spectroscopy.
To determine the luminosities and colors, the individual light 
curves must be extracted from the observed light curve or must 
be inferred from follow-up spectroscopy.
With the known individual masses, the orbital period, and the 
semimajor axis, the orbital elements, e.g., 
$\psi_{\rm P}$, $i$, $\epsilon$, and $\Omega$, 
can be determined from the radial velocity curve, which can be
constructed from  follow-up spectroscopy.
For the determination of the orbital elements, a single spectral
line from either star of the binary will be enough to constrain the
stellar motion provided that the mass ratio $M_2/M_1$ is known
(Smart 1962).
Once all these orbital parameters are known, one can find the projected
separation $\ell$ from equations (2.5) and (2.6).

\section{Degeneracy}
 
There exists a degeneracy in the determination of 
$\hat{r}_{\rm e}$ for case I events.
In general, there are two types of degeneracies in the binary-source
lens geometry.
The first type occurs because the direction of source motion
(with respect to the lens) is not known.
Fortunately, this type of degeneracy does {\it not} affect the
the determination of $\hat{r}_{\rm e}$ because of the radial 
symmetry of Einstein rings.
The other type of degeneracy, which {\it does} affect the
determination of $\hat{r}_{\rm e}$, arises because the $y$-component of
the separation can have two possible values
depending on whether
the sources are located on the opposite
($\Delta\beta_{+} = \beta_1 + \beta_2$) or
at the same  ($\Delta\beta_{-} = |\beta_1 - \beta_2 |$) side
with respect to the lens, resulting in two possible values of 
source-plane Einstein ring size $\hat{r}_{{\rm e}+}$ and 
$\hat{r}_{{\rm e}-}$.
An illustration of two degenerate lens-source positions 
({\it small filled circles}) is shown in panels (a) and (b) 
of Figure 1.

However, the degeneracy can be broken for
binaries with short periods, i.e., $P \lesssim 10t_{\rm e}$.
The principle is simple:\
when the orbital motion is important, the light curves
resulting from the two degenerate lens geometries will be different.
In the panels (a) and (b) of Figure 1, we present tracks
of the actual orbiting source stars
({\it dotted lines}) compared to the hypothetical straight-line tracks
assuming no orbital motion ({\it solid lines}).
The resulting light curves taking the orbital motions into consideration
are shown in the panels (c) and (d).
In the example, 
we assume that the event has $t_{\rm e} = 20\ {\rm days}$ and
the binary is observed to have
$F_{0,1}=F_{0,2}$, $P=100\ {\rm days}$, and inclination angle
$i=0^{\circ}$. 
%and the phase angle $\phi_1 = \pi - \pi_2 = 60^{\circ}$.
For simplicity, we adopt a circular orbit
with a mass ratio ${\cal Q}_{\rm M}=1$, i.e.,
the radii of orbital motion $a_1=a_2=a/2;\ {\rm where}\ a=0.8\ {\rm AU}$
is the semimajor axis,
and the separation of $\ell_j=a_j={\rm constant}$,
corresponding to source mass sum of $ M=M_1+M_2=a^3/P^2=6.82\ M_{\odot}$.
The projected separation between the two source stars in units of 
$\theta_{\rm e}$ is $\hat{\lambda}=0.6$ and 0.2 for the cases 
when source stars are located at the same and opposite sides with 
respect to the lens, as shown in (a) and (b) of Figure 1.
Then there are two possible values of $\hat{r}_{\rm e}\ell/\hat{\lambda}$, 
$\hat{r}_{{\rm e}+} = 1.3$ AU and 
$\hat{r}_{{\rm e}-} = 4.0$AU.
The degeneracy is clearly broken since the light curve (c) is
radically different from (d).

\section{Short-Period Binary Event Light Curves}

We have shown that short-period binaries are important in determining 
the proper motion.
In this section, we derive an analytic form of the short-period 
binary-source event light curve to better understand this type of
event.
If the angular separation between the binary-source stars is 
small compared to the source-lens separation, 
one can treat the difference between the center of light (CL) 
and the position of each source star as a perturbation 
$\delta u_{j}$ of the case when both
stars are at the CL, i.e., 
$$
u_j =
(u^2 - 2u\hat{\lambda}_j\cos \theta_j + \hat{\lambda}_j^2)^{1/2}
\sim u + \delta u_j;
$$
$$
\delta u_j = 
- \hat{\lambda}_j\cos \theta_j
+ {1\over 2 u_{j}} \hat{\lambda}_j^2 
\sin^2 \theta_j,
\eqno(4.1)
$$
where $u=u_{\rm CL}$ and $u_{j}$ are the locations 
of the CL and of each 
source star in units of $r_{\rm e}$, and $\theta_j$ is the angle between 
the lines connecting the CL with the lens and the CL with each source 
(see Fig.\ 2).
Here $\hat{\lambda}_{j}$ are the offsets of the two sources from the CL, 
i.e., $\hat{\lambda}=\hat{\lambda}_{1}+\hat{\lambda}_{2}$ and
$\hat{\lambda}_1 F_{0,1}=\hat{\lambda}_2 F_{0,2}$ and they are
related to the projected physical separations, $\ell_j$, by 
$\hat{\lambda}_j=(\ell_j/r_{\rm e})(D_{\rm ol}/D_{\rm os})= 
\ell_j/\hat{r}_{\rm e}$.
The angles are related by $\theta_2 = \pi -\theta_1$.
The corresponding flux perturbation is then
$$
\delta F_{\rm sep}
\sim A' \sum_{j=1}^2 F_{0,j}\delta u_j  +
{A''\over 2} \sum_{j=1}^2 F_{0,j}\delta u_j^2, 
\eqno(4.2)
$$
where 
$$
A' = {dA \over du}=-{8\over u^2(u^2+4)^{1/2}},\ \
A'' = {d^2A \over du^2}=-{5u^2 + 8 \over u(u^2+4) }A'.
\eqno(4.3)
$$
By combining equations (4.1) and (4.2), and keeping terms up to 
second order in $\hat{\lambda}_j$, one finds the perturbation term to be 
$$
\delta F_{\rm sep}
={A'\over 2u} (F_{0,1}\hat{\lambda}_1^2 + 
F_{0,2}\hat{\lambda}_2^2)
\left[  \sin^2 \theta_1  -
{5u^2 + 8 \over (u^2+4) } \cos^2 \theta_1 \right].
\eqno(4.4)
$$
In the regime where $u\ll 1$, equation (4.4) is approximated by
$$
\delta F_{\rm sep}
\sim{A'\over 2u} (F_{0,1}\hat{\lambda}_1^2 + 
F_{0,2}\hat{\lambda}_2^2)
(3\sin^2 \theta_1 -2),
\eqno(4.5)
$$
resulting in the fractional perturbation of
$$
{\delta F_{\rm sep}\over F}\sim 
-{1\over 2} {3\sin^2 \theta_1 -
2\over {\cal Q}_{\rm L}+2+{\cal Q}_{\rm L}^{-1}}
\left( {\hat{\lambda}\over u}\right)^2
%\propto u^{-2},
\eqno(4.6)
$$
since $A\sim 1/u$ and $A'\sim -1/u^2$ in this regime.
%and $\hat{\lambda}_1=[F_{0,2}/(F_{0,1}+F_{0,2})]\hat{\lambda}$ and 
%$\hat{\lambda}_2=[F_{0,1}/(F_{0,1}+F_{0,2})]\hat{\lambda}$.
Here ${\cal Q}_{\rm L} = F_{0,1}/F_{0,2}$ is the luminosity ratio between 
two stars.
This perturbation alone describes the deviation from the single source 
light curve well enough when
the binary pair is composed of stars of similar type so that 
the location of the CL is close to the the center of mass (CM).

However, the stars are orbiting around the CM
not around the CL.
Therefore, another perturbation term $\delta F_{\rm c}$ arises 
due to the difference between the positions between the CM and CL,
$\delta u_{\rm c}$.
With similar geometry (see Fig.\ 2), the location of the CM is 
$$
u_{\rm CM}=
(u_{\rm CL}^2 
+ 2u_{\rm CL}\hat{\lambda}_{\rm c}\cos\theta_1
+ \hat{\lambda}_{\rm c}^2
)^{1/2}
\sim u_{\rm CL} +  \delta u_{\rm c}; 
$$
$$
\delta u_{\rm c} = \hat{\lambda}_{\rm c}\cos \theta_1 
+ {1\over 2 u_{\rm CL}} \hat{\lambda}_{\rm c}^2 \sin^2 \theta_1 ,
\eqno(4.7)
$$
where the offset between the CL and the CM is 
$$
\hat{\lambda}_{\rm c}=
{{\cal Q}_{\rm L}-{\cal Q}_{\rm M} 
\over (1+{\cal Q}_{\rm M})(1+{\cal Q}_{\rm L})}
\hat{\lambda}.
\eqno(4.8)
$$
Note that 
for the geometry shown in Figure 2, ${\cal Q}_{\rm L} > {\cal Q}_{\rm M}$, 
so $\hat{\lambda}_{\rm c} > 0$.
The flux difference due to the difference in the positions of the CL 
and the CM is then approximated by
$$
\delta F_{\rm c} = (F_{0,1}+F_{0,2})
\left[ A' \delta u_{\rm c} + {A'' \over 2} \delta u_{\rm c}^2 \right]
$$
$$
\sim (F_{0,1}+F_{0,2})
\left[ A'\cos \theta_1 \hat{\lambda}_{\rm c} 
+ {1\over 2u} (A'\sin^2 \theta_1 + A'' u \cos^2 \theta_1)
\hat{\lambda}_{\rm c}^2 \right],
\eqno(4.9)
$$
by keeping terms up to second order in $\hat{\lambda}_{\rm c}$.
For $\hat{\lambda}_{\rm c}\ll u$, the first term always dominates, 
implying a perturbation, 
$$
{\delta F_{\rm c}\over F} \sim
-{A'\cos \theta_{1}\over A} \hat{\lambda}_{\rm c}. 
\eqno(4.10)
$$

One then arrives at the final form of the approximation to the 
total flux, 
$$
F = A(F_{0,1} + F_{0,2}) + \delta F;\ \ \ 
\delta F = \delta F_{\rm c} + \delta F_{\rm sep}.
\eqno(4.11)
$$
The leading first terms of the fractional deviation from a 
single-source event are given by
$$
{\delta F\over F}  \sim 
{({\cal Q}_{\rm L}-{\cal Q}_{\rm M}) \cos \theta_1
\over (1+{\cal Q}_{\rm M})(1+{\cal Q}_{\rm L})}
\left( {\hat{\lambda} \over u} \right)
- 
{1\over 2} {3\sin^2 \theta_1 -2 \over {\cal Q}_{\rm L}+2 + 
{\cal Q}_{\rm L}^{-1} }
\left( {\hat{\lambda} \over u}\right)^2.
\eqno(4.12)
$$
In Figure 3, we present three examples of case II events:
(a) when $\delta F_{\rm sep} \gg \delta F_{\rm c}$ (e.g., pair of stars 
with same mass and luminosity),
(b) when both perturbations take place, and 
(c) when $\delta F_{\rm sep} \ll \delta F_{\rm c}$ 
(e.g., same mass but $F_{0,1} \gg F_{0,2}$ such as a giant-white dwarf pair).
All three example events have $\beta = 0.4$, $t_{\rm e} = 100\ {\rm days}$, 
and orbital period of $P=30\ {\rm days}$, and the orbit is face-on 
and circular, for simplicity.
The masses and luminosities of the individual stars for each case are 
marked in each panel.
The peaks of the curve occur with a frequency of $\sim P/2$ for (a) events
since the varying part of $\delta F_{\rm sep}$ is proportional to  $\sin^2 \theta_1$,
while the frequency is $\sim P$ for (c) events because 
$\delta F_{\rm c}  \propto \cos \theta_1$.
For the general case in which both perturbations play roles, 
the perturbation $\delta F_{\rm c}/F\propto u^{-1}$ decays slowly compared to
$\delta F_{\rm sep}/F\propto u^{-2}$, and thus the $\delta F_{\rm c}$ 
perturbation (period $P$) dominates the wings of the light curve while 
the $\delta F_{\rm sep}$ perturbation 
(period $P/2$) dominates near the peak.

\section{Fraction of Events with Measurable Proper Motion}

What fraction of events have binary sources with periods sufficiently
short to give rise to case II events, and how large are the oscillations
that they generate?  The answer to the first question depends only the source
population, while the answer to the second depends primarily on the positions
of the lenses.  Here, we present rough estimates for these values.
 
For the LMC, about $1/3$ of the sources with $V<19$ are A stars
and $2/3$ are red giants (C.\ Alcock 1993, private communication).  
The typical time scale of observed events is
$t_{\rm e}\sim 37$ days (Pratt et al.\ 1996).
 From
Tables 2 and 3 of Griest \& Hu (1992) about 18\% of A stars have companions
in the period range $1\lesssim P\lesssim 100$ days and for about 2/3 of these
(or 12\% of all A star sources), the mass ratio is $1/{\cal Q}_M>0.35$.  
For simplicity we adopt average parameters $P = 10\ $days, $M=5\ M_{\odot}$,
$1/{\cal Q}_{\rm M}=0.5$, and ${\cal Q}_{\em L}\gg 1$.  
Then the leading factor in equation (4.12) 
$({\cal Q}_{\rm L}-{\cal Q}_{\rm M})/
[(1+{\cal Q}_{\rm M})({\cal Q}_{\rm L})] \sim 1/3$ 
since ${\cal Q}_{\rm L}-{\cal Q}_{\rm M}\rightarrow {\cal Q}_{\rm L}$ 
and $1+{\cal Q}_{\rm L}\rightarrow {\cal Q}_{\rm L}$.
By Kepler's Third Law, the
typical separation is $\ell=0.16\,$AU.  
Then one finds
a typical amplitude of fluctuation
$$
{\delta F\over F}\sim 
{\ell\over 3 u\hat{r}_{\rm e}}
\sim 
{0.1\ {\rm AU}
\over \hat{r}_{\rm e}}
\eqno(5.1)
$$
where we have adopted $u\sim 0.5$ as an average value.  
 
For LMC self-lensing events with a projected speed 
$v\sim 50\ {\rm km\ s}^{-1}$,
we find $\hat{r}_{\rm e} =(D_{\rm os}/D_{\rm ol})v t_{\rm e}
\sim 1\ $AU.  
Hence, the oscillation is of order 10\%.  
Another 30\% of A stars have companions in the range
$0.2< 1/{\cal Q}_{\rm M}<0.35$ and so give rise to effects 
about half this big.
We repeat this calculation for red giants using the binary statistics of
their F star progenitors, but restrict attention to periods 
$10<P<100\ $days since closer companions will be destroyed.  
We find that $\sim 5\%$ of red giants also give rise to oscillations 
of $\sim 10\%$.  
Thus, 8\% of
all source stars should be binaries with periods short enough and companions 
heavy enough to give rise to effects of order 10\%.

For Galactic halo events, the expected source plane Einstein 
ring radius is $\hat{r}_{\rm e} \sim 5r_{\rm e}\sim 22\ {\rm AU}$, 
where we assume that 
$v=200\ {\rm km\ s}^{-1}$ and $D_{\rm ol} =10\ {\rm kpc}$.
Then the expected fluctuation is $\delta F/F \sim 0.5\%$,
which would be difficult to detect, though perhaps not impossible.
However, we stress that 
for the $8\%$ of source stars that are short-period binaries 
even non-detection of the oscillations would be important because it would 
establish that the lens was in the Galaxy and not in the LMC.

Toward the Galactic bulge field, the typical time scale is shorter;
$\sim 20\ {\rm days}$ (Alcock et al.\ 1996c). 
For bulge self-lensing events with $v\sim 200\ {\rm km\ s}^{-1}$, 
a typical source plane Einstein ring radius is 
$\hat{r}_{\rm e} \sim 3\ {\rm AU}$.
For periods of $P\sim 20\ {\rm days}$ and source mass $M\sim 1.5\ M_{\odot}$, 
one finds a separation of $\ell \sim 0.17\ {\rm AU}$.
Typical effects are then $\delta F/F \sim (2/3)(0.17/3)=4\%$ from 
equation (5.1), and thus they are easily
measurable for giants, and possibly for turn off stars as well.

Eclipsing binaries provide excellent opportunities to determine 
$\hat{r}_{\rm e}$, although the expected event rate is very low.
This is because the pure lensing light curve can be easily recovered by 
dividing the observed light curve by the pre-event eclipsing light curve.
In addition, the orbital elements of eclipsing binaries 
can be easily determined due to the known $i=90^{\circ}$.
The eclipse light curve can also be used to help constrain the 
luminosities of the binary sources.
Current experiments have identified many eclipsing stars 
[$\sim {\cal O}(10^4)$] toward the Galactic bulge and LMC 
(Grison et al.\ 1995;  Cook et al.\ 1995;  Ka{\l}u\.zny et al.\ 1995),
but, unfortunately, they are excluded as 
gravitational lensing source stars due to their variability.
Because of the extra information which is automatically available in the 
case of eclipsing binaries, they should be reincluded in the lensing search.
Luckily enough,
there already has been reported a candidate event (EROS2, Ansari et al.\ 1995)
with a microlensing-type light curve superimposed on top of periodic eclipsing 
binary variability.

\section{Identification of Lens Population}

Once the values of $\hat{r}_{\rm e}$ and thus $\hat{v}$ are known,
one can strongly constrain the nature of individual lenses.
First, for the known distance to source stars, e.g., 
$d_{\rm bulge}= 8\ {\rm kpc}$
and $d_{\rm LMC} = 50\ {\rm kpc}$ 
toward the Galactic bulge and LMC, respectively,
determining $\hat{r}_{\rm e}$ is 
equivalent to measuring the proper motion
$\mu = \theta_{\rm e}/t_{\rm e}$.
While the time scale is a function of three parameters,
$t_{\rm e} = t_{\rm e} (M_{L}, v, D_{\rm ol})$, 
the Einstein ring is a function of only two,
$\theta_{\rm e}=\theta_{\rm e} (M_L,D_{\rm ol})$, and hence provides
less degenerate information.
Secondly, once $\mu$ is measured, one can easily distinguish 
Galactic halo from LMC self-lensing events from the difference in $\mu$
because $\mu_{\rm LMC} \ll \mu_{\rm halo}$.
In Table 1, we present the expected typical values of
$\langle \hat{r}_{\rm e} \rangle$ and $\langle \hat{v}\rangle$
for the LMC disk, LMC halo, and Galactic halo events.
Also presented are the typical parameters for the geometry,
$\langle D_{\rm ol}/D_{\rm os}\rangle$,
the lens mass $\langle M_L \rangle$, and
the transverse speed, $\langle v \rangle$, which are used
in determining the expected
$\langle \hat{r}_{\rm e} \rangle$ and $\langle v \rangle$.
We adopt heavier masses for LMC disk lenses because the lenses are expected
to be objects above hydrogen-burning limit.
As a tool to identify the lens population, $\hat{v}$ will be more
useful than $\hat{r}_{\rm e}$ because
$\hat{v}$ does not depend on the lens mass,
which, for any given event, is completely unknown.
However, we note that for observations toward the Galactic bulge 
it is difficult to distinguish Galactic 
disk from Galactic bulge lenses using $\mu$ since both have similar 
distributions (Han \& Gould 1995).

\section{An Immediate Application}

Although the events detected to date toward the LMC bear 
no obvious signatures of binary sources, it would nonetheless be 
interesting to check the sources for binarity.
If any of the sources are binaries that would produce effects for
the case of an LMC self-lensing event,
the failure to detect an effect would prove that the lens was Galactic.
To illustrate the nature of the required observations, 
suppose that one of the sources were an A-type binary with 
$M=5\ M_{\odot}$, ${\cal Q}_{\rm M}=2$, ${\cal Q}_{\rm L}=16$, 
$P=100\ {\rm days}$, and a circular orbit.
The velocity oscillations of the primary would then be sinusoidal with 
a full width $\Delta v = 50\ {\rm km\ s}^{-1}\sin i$.
Hence, for all but the rare, nearly face-on orbits,
the velocity variations 
would show up in 3 or 4 1-2 hr observations on a 4\ m telescope 
each separated $\sim 1$ month.
Note that the velocity can be measured with an uncertainty of 
$\lesssim 10\ {\rm km\ s}^{-1}$ for 19 mag A-type and K giant stars
(1996 C.\ Pryor, private communication).
Other parameters considered in this paper ($P<100\ {\rm days}$, 
${\cal Q}_{\rm M} < 2$) would produce even stronger velocity variations.
Once this source was  identified as a binary candidate, it could be 
subjected to more intensive observation which would reveal the spectral 
class (and hence mass $M_1=3.3\ M_{\odot}$) of the primary as well 
as the circular character of the orbit.
Unless the observations were very intensive, the spectral signature 
of the secondary (F star) might  not be detected (single-line binary).
For definiteness we assume that this lack of detection can be interpreted
as an upper limit on its mass, $M_2<2\ M_{\odot}$.
Suppose for example $i=45^{\circ}$, so that 
$\Delta v = 35\ {\rm km\ s}^{-1}$.
Without detection of the secondary, this inclination would be unknown, 
but could be constrained to be $0.63 < \sin i < 1.0$ based on the mass 
ratio limit ${\cal Q}_{\rm M} > 1.65$.
Hence the semimajor axis of the orbit of the primary would lie in the 
range $0.5\ {\rm AU} < a_1 < 0.8\ {\rm AU}$.
Since this entire range is of the same order as $\hat{r}_{\rm e}$
for a typical LMC self-lensing event,  
there would be a dramatic effect on the light curve if the event were 
in fact of this type.
Thus, in this case, one would determine that the event was Galactic 
even without detection of the secondary.
This is in fact the case for most of the parameter space of case II events.

In brief, the observations should follow three stages.
First, sparse monitoring to determine which sources are binaries.
Second, more intensive monitoring to determine the orbital elements
and so possibly distinguish between Galactic and LMC events.
Third, highly intensive follow up to try to detect the secondary 
if the questions about the event warrant this effort.

\acknowledgments
We would like to thank G.\ Newsom and C.\ Pryor for making
useful comments and the refree for helpful suggestions.
This work was supported by a grant AST 94-20746 from the NSF.

\newpage
\bigskip
\begin{center}
\begin{tabular}{lcccrrr}
\hline
\hline
\multicolumn{1}{c}{population} &
\multicolumn{1}{c}{$\langle D_{\rm ol}/D_{\rm os}\rangle$} &
\multicolumn{1}{c}{$\langle M_L \rangle$} &
\multicolumn{1}{c}{$\langle r_{\rm e} \rangle$} &
\multicolumn{1}{c}{$\langle \hat{r}_{\rm e} \rangle$} &
\multicolumn{1}{c}{$\langle v \rangle$} &
\multicolumn{1}{c}{$\langle \hat{v} \rangle$} \\
\multicolumn{1}{c}{} &
\multicolumn{1}{c}{} &
\multicolumn{1}{c}{$(M_{\odot})$} &
\multicolumn{1}{c}{(AU)} &
\multicolumn{1}{c}{(AU)} &
\multicolumn{1}{c}{$({\rm km}\ {\rm s}^{-1})$} &
\multicolumn{1}{c}{$({\rm km}\ {\rm s}^{-1})$} \\
\hline
LMC disk & 0.99 & 0.3 & 1.05 & 1.06 & \ 50 & \ \ 50 \\
LMC halo & 0.94 & 0.1 & 1.44 & 1.53 & 100 & \ 106 \\
Galactic Halo & 0.20 & 0.1 & 2.40 & 12.0\  & 220 & 1100 \\
\hline
\end{tabular}
\end{center}
 
\noindent
{\footnotesize
{\bf Table 1}:\ Characteristic values of $\hat{r}_{\rm e}$
and $\hat{v}$ for different populations of lenses.
Also presented are the typical parameters for the geometry,
$\langle D_{\rm ol}/D_{\rm os}\rangle$,
the lens mass $\langle M_L \rangle$, and
the transverse speed, $\langle v \rangle$, which are used
in determining the expected
$\langle \hat{r}_{\rm e} \rangle$ and $\langle v \rangle$.
}
\bigskip

%-------------------------------------------------------------------------
\newpage
\centerline{\bf FIGURE CAPTION}
\bigskip

\noindent
{\bf Figure 1}:
Breaking the degeneracy in $\hat{r}_{\rm e}$.
Without orbital motion, the lens geometries (a) and (b)
would produce the same light curves.
However, when the orbital motion is important, the light curves
resulting from the two degenerate lens geometries will be different
as shown in the panels (c) and (d).
The light curves of the primary and secondary stars are presented 
by {\it thin solid and dotted lines}, while the sum of flux from both 
the primary and secondary is shown with {\it thick solid lines}.
The tracks of the actual orbiting source stars
({\it dotted lines}) are compared to
the straight-line tracks ({\it solid lines}) assuming no motion.

\noindent
{\bf Figure 2}:
Case II binary-source geometry.
Since $u\gg \hat{\lambda}_j$, the separations
$\delta u_j \sim \left\vert {\bf u}_{\rm CL}-{\bf u}_j \right\vert$
can be treated as a perturbation of the case when both
stars are at the center of light (CL).
The source stars are denoted by filled circles
(bigger for the primary) and the lens is marked by `L'.
Similarly, one can treat the separation
$\delta u_{\rm c}\sim \left\vert {\bf u}_{\rm CM}-
{\bf u}_{\rm CL} \right\vert$
as a perturbation to take account of the motion of the CL around 
the center of mass (CM).

\noindent
{\bf Figure 3}:
Case II event light curves:
(a) when $\delta F_{\rm sep} \gg \delta F_{\rm c}$ (e.g., pair of stars
with same mass and luminosity),
(b) when both perturbations take place, and
(c) when $\delta F_{\rm sep} \ll \delta F_{\rm c}$
(e.g., same mass but $F_{0,1} \gg F_{0,2}$ such as a giant-white dwarf pair).
All three example events have $\beta = 0.4$, $t_{\rm e} = 100\ {\rm days}$,
and orbital period of $P=30\ {\rm days}$.
The mass (in $M_{\odot}$) and luminosity of individual stars for each case
are marked in each panel.
The peaks of the curve occur with a frequency of $P/2$ for (a) events
since $\delta F_{\rm sep} \propto \sin^2 \theta_1$,
while the frequency is $P$ for (c) events because
$\delta F_{\rm c}  \propto \cos \theta_1$.
For the general case in which both perturbations play roles,
the perturbation $\delta F_{\rm c}/F\propto u^{-1}$ decays slowly compared to
$\delta F_{\rm sep}/F\propto u^{-2}$, and thus the $\delta F_{\rm c}$
perturbation (period $P$) dominates the wings of the light curve.

\newpage
\postscript{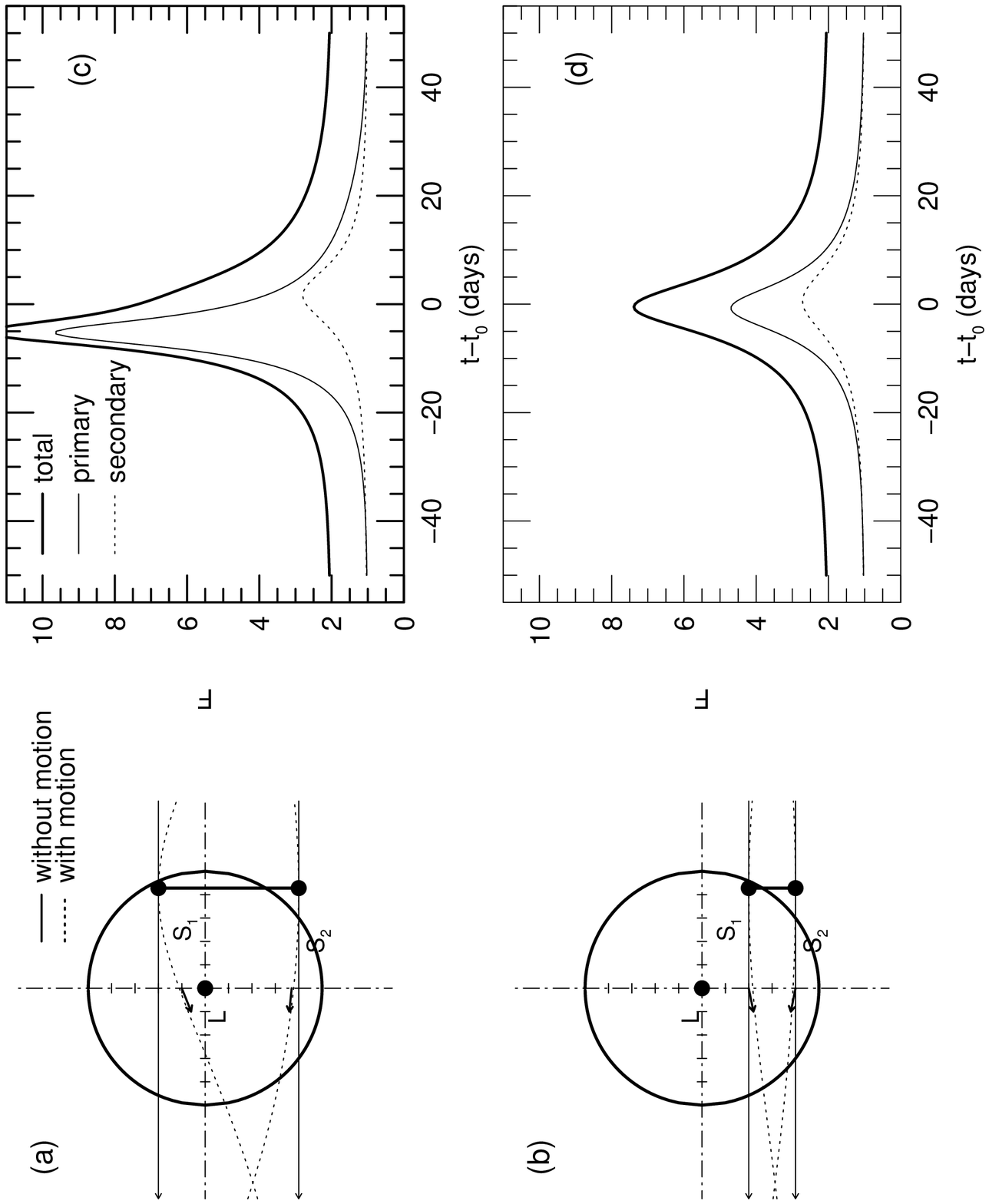}{1.3}

\newpage
\postscript{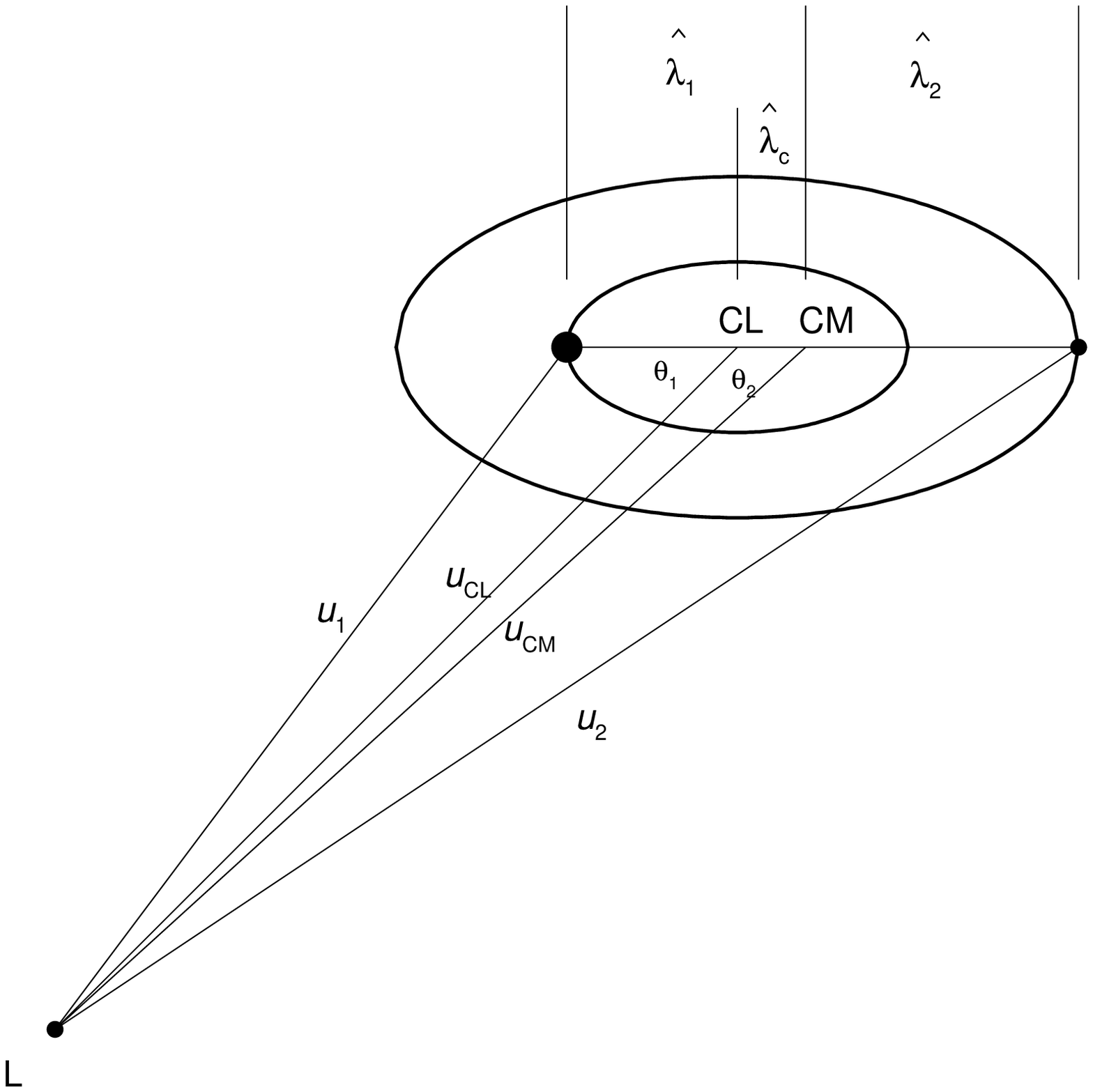}{1.0}

\newpage
\postscript{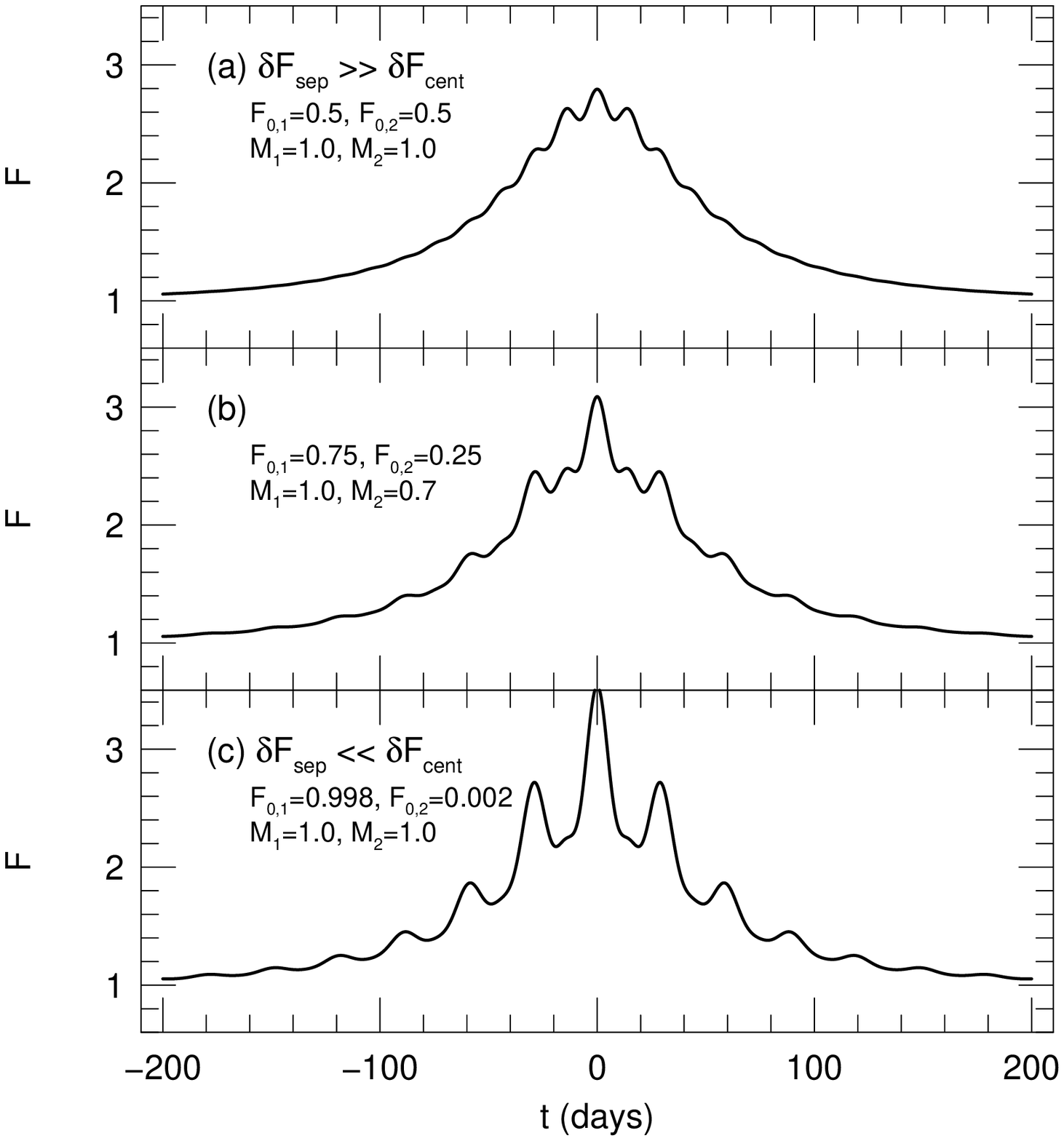}{1.1}

\end{document}